\documentstyle[12pt,russian]{article}
\begin{document}
\newcommand{\bc}{\begin{center}}
\newcommand{\ec}{\end{center}}
\newcommand{\tg}{\mathop{\rm tg}\nolimits}
\renewcommand{\refname}{}
\def\d{\partial}
\def\dif{\stackrel{\leftrightarrow}{\d}}
\def\pint{\int\limits_{0}^{\infty}{\kern-.81em\hbox{{\sf --}}}~~~}
\def\lss{\mathop{<}\limits_{\sim}}
\newcommand{\con}{{\large{\sf D}}\hspace{-2.1ex}\raisebox{2.5pt}{\tiny
$\vee$}\hspace{.3ex}\raisebox{2.5pt}{\tiny $\wedge$}}
\makeatletter
\renewcommand{\@makefnmark}{\hbox{\mathsurround=0pt$^{\@thefnmark{})}$}}
\renewcommand{\@makefntext}[1]{\parindent=1em\noindent
\hbox to 1.8em{\hss$^{\@thefnmark{})}$}#1}
\makeatother

\def\<{\langle}
\def\>{\rangle}
\def\[{\left[}
\def\]{\right]}
\def\({\left(}
\def\){\right)}
\def\{{\left\lbrace}
\def\}{\right\rbrace}
\def\ds{\displaystyle}
\def\ts{\textstyle}
\def\ss{\scriptstyle}
\def\wt{\widetilde}
\def\sh{\mathop{\rm sh}\nolimits}
\def\ch{\mathop{\rm ch}\nolimits}
\def\th{\mathop{\rm th}\nolimits}
\def\diag{\mathop{\rm diag}\nolimits}
\def\ln{\mathop{\rm ln}\nolimits}
\bc
\setcounter{page}{1}
{\bf {\Large
Vacuum Energy and Casimir Force in a Presence of Skin-depth Dependent
Boundary Condition\\}}
\vspace{2ex}
\large

S.L.~Lebedev \footnote { E-mail: lsl@chuvsu.ru
Permanent adress: 428000 Cheboksary, Chuvash I.Ya.~Yakovlev State Pedagogical
University.}\\

\ec
\rm
\begin{quote}
\small

{The vacuum energy-momentum tensor (EMT) and the vacuum energy
corresponding to massive scalar field on
$\Re_{t}\times [0,l] \times \Re^{D-2}$ space-time with boundary condition
involving a dimensional parameter ($\delta$) are found. The dependent on the
cavity size $l$ Casimir energy $\wt E_{C}$ is a uniquely determinable function
of mass $m$, size $l$ and "skin-depth" $\delta$. This energy includes the "bulk"
and the surface (potential energy) contributions. The latter dominates
when $l \sim \delta$. Taking the surface potential energy into account
is crucial for the coincidence between the derivative $-\d \wt E_{C}/\d l$ and
the $ll$-component of the vacuum EMT. Casimir energy $\wt E_C$ and the bulk
contribution to it are interconnected through Legendre transformation, in which
the quantity $\delta^{-1}$ is conjugate to the vacuum surface energy multiplied
by  $\delta$. The surface singularities of the vacuum EMT
do not depend on $l$ and, for even $D$, $\delta =0$ or
$\infty$, possess finite interpretation. The corresponding vacuum energy is
finite and retains known analytical dependence on the dimension $D$.} \\
\end{quote}

\bc {\bf 1. Introduction} \ec

The determination of vacuum energy of fields confined to finite volumes is the
basic part in the calculations of Casimir force as well as of some bag
characteristics of hadrons \cite{MT0,HK}. At the same time the renormalization
approach appears to be not unique: the counterterms necessary to make the
vacuum energy finite can depend on the cavity size so that the regularization
dependence of Casimir energy appears \cite{BVW,CVZ}. Concrete calculations for
the scalar case and Dirichlet or Neumann boundary condition on a spherical
boundary have been performed in the recent papers \cite{LR,BEKL,NP}. General
consideration \cite{DC} shows however that at least for these boundary
conditions the surface singularities of the vacuum EMT (and, hence, the
aforementioned counterterms) can not depend on the size of the cavity. The
detection of such dependences in the papers listed above is a consequence of a
coincidence between the local (curvature) and global (size) parameters
determining the spherical boundary.

We consider the vacuum characteristics of massive scalar field defined on the
domain $\Re_{t}\times [0,l] \times \Re^{D-2}$. In the absence of curvature we
introduce "skin-depth" parameter $\delta$ \cite{MT5} in boundary condition
\footnote{We use the system of units where $\hbar=c=1$.
The signature of metric is $(+,-,-,\ldots)$. $x^{\mu}=(t,x_{1},\bf{x}_{\perp})$.
Brackets $[ , ]_{+}$ mean anticommutator.}
$$
\d_1\varphi(t,0,{\bf x}_{\perp})=\frac{1}{\delta}\varphi(t,0,{\bf x}_{\perp}),
\quad \varphi(t,l,{\bf x}_{\perp})=0
\eqno{(1)}
$$
This boundary condition implies the  presence of
the surface potential energy \cite{lsl}
$$
\frac{1}{2\delta}\int d^{D-2}{\bf x_{\perp}} \;\varphi^2(t,0,{\bf
x_{\perp}})
\eqno{(2)}
$$
in the total hamiltonian of the model. The latter is, as usual, space integral
of 00-component of the energy-momentum tensor
$$ \wt T_{\mu\nu}=\frac12\[\d_{\mu}\varphi ,
\d_{\nu}\varphi\]_{+}-\frac12 g_{\mu\nu}(\d\varphi)^{2}+\frac12 g_{\mu\nu}
m^{2}\varphi^{2}+\xi\d^{\lambda}(g_{\mu\nu} \[\varphi,\d_{\lambda}\varphi\]_{+}-
g_{\mu\lambda}\[\varphi,\d_{\nu}\varphi\]_{+})
\eqno{(3)}
$$
where the last divergence term should be taken with coefficient $\xi=1/4$. This
is the only value of $\xi$ guaranteeing the diagonality of hamiltonian and its
conservation \cite{lsl}.

\bc
{\bf 2. Energy-momentum tensor and energy of vacuum}
\ec

The normalized solutions of Klein-Gordon-Fock equation have the form:

$$
\wt \varphi_{k}(x)=(2\omega)^{-\frac12}(2\pi)^{1-\frac{D}{2}}\exp{[-\it
i\omega t+\it i \bf{qx_{\perp}}]}\psi_{k}(x_{1}),
\eqno{(4)}
$$
where $\omega=\sqrt{m^{2}+{\bf q^{2}}+k^{2}}$,
and the descrete set of functions
$$
\psi_{k}(x)=N_k \sin{k(x-l)}, \quad
N_k=\[\frac{l}{2}\(1-\frac{\sin{2kl}}{2kl}\)\]^{-1/2}
 \eqno{(5)}
$$
represents the resonator modes. Wave number
$k=z/\delta \quad(\wt l \equiv l/\delta)$ should be determined from the spectral
equation
$$
\Delta(z)\equiv z^{-1}\tg{z\wt l}+1=0,
\eqno{(6)}
$$
following from boundary condition (1). The basis of functions (4) is used to
define creation-annihilation operators and vacuum state in a standard manner
\cite{MT0,BD} \footnote{It should be noted, however, that canonical
quantization of the system (3) for $\xi =1/4$ as a system with a higher
derivatives, should account for its degeneracy \cite{lsl}.}.

With the help of Cauchy theorem applied to meromorphic function $\Delta(z)$, it
is possible to transform the sums over transcendential roots of eq.(6) into
corresponding integrals. The scheme of calculation fits in the formula
$$
\sum_{z_{n\delta}>0}\,\frac{f(z_{n\delta})}{\ds 1-\frac{\sin
{2z_{n\delta}\wt l}}{2z_{n\delta}\wt l}} =
\frac{\wt l}{\pi}\,\int\limits_{0}^{\infty}\,f(z)\,dz-\frac{\wt l}{2\pi i}
\int\limits_{0}^{\infty}\,\frac{(1-t)[f(it)-f(-it)]}{\sh{t\wt
l}+t\ch{t\wt l}}\,e^{-t\wt l}\, dt,
\eqno{(7)}
$$
where $f(z)$ is a function analitical in the right half-plane, and
$z_{n\delta}\, (n=1,2,\cdots)$ are the (real) roots of eq.(6). With the use of
eq.(7) one could attach the integral representation to the unrenormalized vacuum
EMT. The renormalization is reduced to the subtraction of Minkowskian space
contribution (D-regularization \cite{col} supposed):
$$
\<T_{\mu\nu}\>_{M}=\frac{\Gamma(\frac{1-D}{2})}
{\sqrt{\pi}(4\pi)^{D/2}}\int_{0}^{\infty}\;dk\;M^{D-3}
\diag\[\begin{array}{c}-M^2\\
k^2(1-D)\\ M^2
\end{array}\] ,
\eqno{(8)}
$$
$M\equiv (m^2 + k^2)^{1/2}$. The final answer i.e. renormalized EMT of vacuum,
looks like
$$
\<\wt T_{\mu\nu}\>_{\delta,l}={{\rm K}_{D}\over \delta}
\int\limits_{\mu}^{\infty} dt(t^2-\mu^2)^{\frac{D-1}{2}}
 \{ \ds{{(1-t) e^{-t\wt l}\over\sh{t\wt
l}+t\ch{t\wt l}}} \; \diag\[\begin{array}{c} \ch{2t x'}-1 \\
\ds{\frac{t^{2}(1-D)}{t^2-\mu^2}} \\1-\ch{2t x'} \end{array} \] +
e^{-2t x'} \diag\[\begin{array}{c} 1 \\ 0 \\ -I \end{array}\]  \}.
\eqno{(9)}.
$$
$(D-2)$-dimensional diagonal of $\<\wt T_{\mu\nu}\>_{\delta,l}$ corresponding to
${\bf x_{\perp}}$ contains expressions of the same type. The following
notations were used in eq.(9):
$$
{\rm K}_{D}=\frac{\delta^{1-D}}{2(4\pi)^{\frac{D-1}{2}}\Gamma\(\frac{1+D}{2}\)},
\;\mu= m\delta,~  x'=(l-x_1)/\delta .
\eqno{(10)}
$$

The energy of vacuum could be determined directly, without resort to eq.(9), as
a sum of half-frequences interpreted e.g. with the help of zeta-regularization
method \cite{BD}. At the same time, Green function method \cite{MT0,DC,BD} used
to obtain vacuum energy should rely on the tensor $\wt T_{\mu\nu}$ (3)
\cite{lsl}, but not its first three terms
$$
T_{\mu\nu}=\frac12\[\d_{\mu}\varphi ,
\d_{\nu}\varphi\]_{+}-\frac12 g_{\mu\nu}(\d\varphi)^{2}+\frac12 g_{\mu\nu}
m^{2}\varphi^{2}.
\eqno{(11)}
$$
$11$-component of tensor (3) gives Casimir pressure
$P=\<\wt T_{11}\>_{\delta,l}$ which, because of translational symmetry,
coincides with $\< T_{11}\>_{\delta,l}$ (i.e. the divergence term in (3) does
not affect the pressure). The role of divergence term is evident from the fact
that value $\xi=1/4$ is the only one at which equality
$$
-\frac{\d \wt E_{vac}}{\d l}= \< \wt T_{11}\>_{\delta,l}
\eqno{(12)}
$$
holds. Notice that conservation of the total energy \cite{lsl} and equality (12)
both rely on the value $\xi=1/4$.

In proving the eq.(12) it is essential that vacuum energy per unit area of the
boundary
$$
\wt E_{vac}=
\int_{x_0}^{l-y_{0}}dx_1\;\<\wt T_{00}\>_{\delta,l}= \wt E_{C}(l,\delta)+\wt
E_{w1}(x_0,\delta)+ \wt E_{w2}(y_0,0)+\cdots ,
\eqno{(13)}
$$
where dots denote terms vanishing at $x_0,y_0\rightarrow 0$ as well as at
$l\rightarrow \infty$. Only finite part of $\wt E_{vac}$, i.e. Casimir energy
$$
\wt E_{C}(l,\delta)={\rm K}_{D}\int\limits_{\mu}^{\infty} dt\;
\ds{{(t^2-\mu^2)^{\frac{D-1}{2}}
e^{-t\tilde l}\over\sh{t\wt l}+t\ch{t\wt l}}}\;\[\wt l(t-1)-\frac{1}{t+1}\]\, ,
\eqno{(14)}
$$
depends on the size of the domain and vanishes when $l\rightarrow \infty$.
Boundary divergences are present in r.h.s. of eq.(13) in the form of the
energies of the walls, $\wt E_{w1,2}$. For example,
$$
\wt E_{w1}(x_0,\delta)={\rm K}_{D}\mu^{D-1}\int\limits_1^\infty
\ds{\frac{d\xi}{2\xi}}\;
(\xi^2-1)^{\frac{D-1}{2}}\;\ds{\frac{1-\mu\xi}{1+\mu\xi}}\; e^{-2mx_0\xi}.
\eqno{(15)}
$$
The specific representation of energies $\wt E_{w1,2}$ depends on regularization
employed, but Casimir energy is independent of the latter. Notice that the terms
discarded in the r.h.s. of eq.(13) have the property of "double vanishing" (with
respect to limits of $x_0,y_0$ and $l$) so that energies $\wt E_{w1,2}$ of the
walls are determined uniquely {\it within} the given regularization scheme.

By a complete analogy with $\wt E_{vac}$, one can find its bulk ($ E_{vac}$)
and surface ($\Pi_{vac}=\wt E_{vac}- E_{vac}$) components, the first
dertermined with the help of density $T_{00}$ (11). For either of these
components the expansion of the form (13) exists giving rize to
$E_{C}(l,\delta)$ and $\Pi_{C}(l,\delta)$ correspondingly. For example,

$$
\Pi_{C}(l,\delta)={\rm K}_{D}(1-D)\;\int\limits_{\mu}^{\infty} dt\;
 \ds {{t^{2}(t^2-\mu^2)^{\frac{D-3}{2}} {e^{-t\tilde l}}\over {(t+1)
(\sh{t\wt l}+t\ch{t\wt l})}}},
\eqno{(16)}
$$
and  $E_{C} =\wt E_{C} - \Pi_{C} $. Since $\Pi_{C}(l,\delta)\neq 0$, derivative
$-\d E_{C}/\d l$ does not coincide with the Casimir pressure (12). Nevertheless,
$$
\delta \cdot \frac{\d \wt E_{C}}{\d \delta}=- \Pi_{C} ,
\eqno{(17)}
$$
and, with the notations $\lambda =\delta^{-1}, f=\Pi_{C}/\lambda$, one finds
that $\wt E_{C}$ and $E_C$ are interconnected through Legendre transformation:
$$
\wt E_C(l,\lambda)=E_C(l,f(l,\lambda))+\lambda f(l,\lambda),
\eqno{(18)}
$$
so that
$$
\( \frac{\d E_C}{\d l}\)_{f}=\( \frac{\d \wt E_C}{\d l}\)_{\lambda},\,\, {\rm
and}\, \(\frac{\d E_C}{\d f}\)_l =-\lambda.
\eqno{(19)}
$$
Now energy $E_C$ determines the Casimir pressure but under specific condition of
constancy of the quantity $f$ being one half the "Casimir" part of
$\< \varphi(0)^2 \>$, see (2).

\bc {\bf 3. Asymptotic properties} \ec

For the purpose of comparison with electrodynamics, here we consider a massless
case. The behaviour of integral (14) in Dirichlet ($\delta \ll l$) and Neumann
($\delta \gg l$) regimes is displayed by the following expansions ($m=0$):
$$
\wt E_{C}(l,\delta) =
{\rm K}_{D}\;\ds\frac{\zeta(D)\Gamma(D)}{(2\wt l)^{D-1}}
\[-1+C^{1}_{D-1}\; \wt l^{-1}-
C^{2}_{D}\; \wt l^{-2}+
C^{3}_{D+1} \(1+\ds\frac{\zeta(D+2)}{2\zeta(D)}\)\;\wt l^{-3}\],
\eqno{(20)}
$$
$$
\wt E_{C}(l,\delta)={\rm K}_{D}\wt l^{-D+1}
\{\begin{array}{l}A_{D}- (D-1)A_{D-2}\wt l,\qquad D=3,4,\dots,
\\  \\
A_2+\wt l \ln(4\gamma_{E}\wt l/\pi)-\wt l,\qquad \qquad D=2. \end{array}
\right.
\eqno{(21)}
$$
Here  $C^{m}_{n}$ denotes binomial coefficient,
$A_{D}=2^{1-D}(1-2^{1-D})\Gamma(D)\zeta(D)$, $\ln\gamma_{E}=0.577$, and eq.(21)
contains the leading corrections only. The expansion of pressure
$\<\wt T_{11}\>_{\delta,l}$ corresponding to (20) has the form ($\delta\ll l)$:
$$
\begin{array}{c}{\ds\<\wt T_{11}\>_{\delta,l} =\frac{(D-1)\Gamma(D)\zeta(D)}
{(4\pi)^{\frac{D-1}{2}}\Gamma\(\frac{1+D}{2}\)}\;(2l)^{-D}\times} \\  \\
\times \[-1+C^{1}_{D} \wt l^{-1}-C^{2}_{D+1} \wt l^{-2}+
C^{3}_{D+2} \(1+\ds\frac{\zeta(D+2)}{2\zeta(D)}\)\;
\wt l^{-3}+\ldots\], \end{array}
\eqno{(22)}
$$
and for Neumann regime (21)
$$
\<\wt T_{11}\>_{\delta,l}=\frac{(D-1)~l^{-D}}
{2(4\pi)^{\frac{D-1}{2}}\Gamma\(\frac{D+1}{2}\)}\;
\[A_{D} - (D-2)A_{D-2}~\wt l+\ldots \].
\eqno{(23)}
$$
The four terms in the square brackets of eq.(22) (term with $\zeta$-functions
excluding) could be obtained through the Taylor series for $(l+\delta)^{-D}$.
Thus, eq.(22) demonstrates the role of $\delta$ as penetration depth
\cite{MT5}.  Unlike Dirichlet case (22), the correction term in the
brackets of eq.(23) emerges due to surface energy $\Pi_C$ only. Numerical
analysis of the formulas (14) and (16) taken at $m=0$ shows the dominance of the
surface contribution $\Pi_C$ over the bulk one ($E_C$) in the region $l\sim
(1\div 4)\delta$. The difference in signs between those quantities is
responsible for the shift of the position of the minimum from $l\sim 4.5\delta$
(for $E_C$) to $l\sim 1.8\delta$ (for $\wt E_C$). At that time, the qualitative
behaviours of energies $E_C$ and $\wt E_C$ as functions of $l$ are alike
showing a typical van-der-vaals character.

At $D=4$ formula (22) takes the form
$$
\<\wt T_{11}\>_{\delta,l}=\frac{\pi^2}{480 l^4}
\[-1+4\frac{\delta}{l}-10\frac{\delta^2}{l^2}+20\(1+\frac{9\pi^2}{185}\)
\frac{\delta^3}{l^3}+\cdots \],
\eqno{(24)}
$$
and should be compared with the corresponding expression in the case of
electromagnetic field confined to the region between impedance walls
($\delta$- skin-depth parameter):
$$
P=\frac{\pi^2}{240
l^4}\[-1+\frac{16}{3}\frac{\delta}{l}-24
\frac{\delta^2}{l^2}+\frac{640}{7}\(1+\frac{9\pi^2}{740}\)
\frac{\delta^3}{l^3}+...\].
\eqno{(25)}
$$
Formula (25) except the last term $\sim \delta^3 / l^3$ in the square brackets,
was extracted from \cite{MT0,MT5}. The corresponding coefficients of
expansions (25) and (24) are related to each other in ratios 1, 1.33, 2.4,
3.46, hence showing the growing effect of spin when going deeper into the
boundary.

\bc {\bf 4. Interpretation of surface singularities at $\delta=0$ and $\infty$}
\ec
The surface singularities of the vacuum EMT is a stumbling-block problem for any
field theory expected to establish the connection between its local properties
and observational predictions \cite{DC}. It is shown below that the method of
dimensional regularization could be applied not only to interpret the singular
sums like $E_{vac}=\sum\limits_{\nu}\;\frac{1}{2}\omega_{\nu}$, but gives a
reasonable finite answer for the energy density of vacuum as well. The method
works for even $D, \delta=+0$ \footnote{The singular nature of Dirichlet limit
(contrary to Neumann one) is explained in \cite{lsl}.} or $\infty$ and leads
to an energy expression consistent with the one of $\zeta$-regularization
method.

Boundary divergences are represented in r.h.s. of eq.(13) by the energies of the
walls ($\wt E_{w1,2}$). Correspondingly, we consider 00-component of a tensor
(9) in the limit $l\rightarrow \infty, \, \delta=+0$ (alternative case
$\delta=\infty$ differs from (26) only in sign):
$$
\<\wt T_{00}\>_{0,\infty} = {\ds \(\frac{m^2}{2\pi}\)^{\ts{D\over 2}}}\;
\{\begin{array}{c}
\rho^{-D/2}\;K_{D/2}(\rho),\qquad 0<\rho\lss 1, \; x/l \rightarrow 0; \\ \\
(\rho ')^{-D/2}\;K_{D/2}(\rho '),\qquad 0<\rho '\lss 1,\; x'/l
\rightarrow 0   \end{array} \right.
\eqno{(26a,b)}
$$
($\rho=2mx, \rho '=2mx'=2m(l-x)$). Exploiting recurrence relations between
modified Bessel functions $K_{\nu}(\rho)$ \cite{BE2}, let us transform
energy density (26a):
$$
\<\wt T_{00}\>_{0,\infty}\rightarrow
\<\wt T_{00}\>_{0,\infty}- \frac{d}{d\rho}f_{D}(\rho)=\( \frac{m^2}{2\pi}\)^
{\ts {D\over 2}}
\, \frac{\rho^{1-\epsilon}\;K_{\epsilon-1}(\rho)}{(1-D)(3-D)\cdots
(1-2\epsilon)}.
\eqno{(27)}
$$
Here $\epsilon \equiv \frac{D}{2}-n\rightarrow 0,\, n=1,2,\dots$, and function
$$
f_{D}(\rho)=\( \frac{m^2}{2\pi}\)^{\ts {D\over 2}}\[ \frac{\rho^{1-{D\over
 2}}K_{{D\over 2}}(\rho)}{1-D}+ \frac{\rho^{2-{D\over 2}}K_{{D\over
 2}-1}(\rho)}{(1-D)(3-D)}+ \dots +
 \frac{\rho^{1-\epsilon}K_{\epsilon}(\rho)}{(1-D)(3-D)\cdots (1-2\epsilon)}\]
\eqno{(28)}
$$
possesses following properties: $i)$ for $\Re{\rm e}D <0$
($\Re {\rm e}~ \epsilon <-n$)

$$
f_{D}(0)=f_{D}(\infty)=0,
\eqno{(29)}
$$
so that divergence addition in the l.h.s. of eq.(27) does not affect the vacuum
energy of the wall \cite{BEKL,lsl,AW}
$$
\wt E_{w1}(0,0)=\(\frac{m^2}{2\pi}\)^{\ts{D\over 2}}\;
\int_{0}^{\infty}dx\;\rho^{-D/2}\;K_{D/2}(\rho)=
 \frac{m^{D-1}\Gamma\(\frac{1-D}{2}\)}{8(4\pi)^{\frac{D-1}{2}}}.
\eqno{(30)}
$$
(the vacuum energy of Neumann wall $\wt E_{w1}(0,\infty)=-\wt E_{w1}(0,0)
=-E_{w1}(0,0)$);

$ii)$ a term $f_{D}^{'}(\rho)$ in the l.h.s. of eq.(27) taken at "physical"
values of $D=2,4,6,\cdots$, acts as a counterterm eliminating all nonintegrable
singularities of density (26a);

$iii)$ this counterterm preserves an exponential decreasing behaviour of the
modified density (27) at $x>\hbar/mc$.
The latter is not still uniquely defined: one can add to $f_{D}(\rho)$ any
regular function having the property (29) (see \cite{DC,BD}).

Thus, at $\delta=+0$ the total energy of vacuum includes Casimir part \cite{AW}
$$
\wt
E_{C}(l,0)=-\frac{m^{D}l}{(4\pi)^{\frac{D-1}{2}}\Gamma(\frac{D+1}{2})}
\int\limits_{1}^{\infty}\,d\xi\,\frac{(\xi^2-1)^{\frac{D-1}
{2}}}{e^{2\xi ml}-1}
\eqno{(31)}
$$
and doubled topological (according to \cite{MT0}) energy (30) caused by the
presence of the edge of the manifold. At $\delta=\infty$ the total energy of
vacuum reduces to $\wt E_{C}(l,\infty)$ because the energies of Neumann ($x=0$)
and Dirichlet ($x=l$) walls cancell out. Energy $\wt E_{C}(l,\infty)=
E_{C}(l,\infty)$. The integral representation for it can be obtained from (31)
by substitution "$+$" for "$-$" in the denominator of the integrand and
by multiplying the whole expression by $(-1)$.

Dimensional regularization method does not result in satisfactory expressions
for the vacuum energy when $0<\delta<\infty$. As it may be seen from eq.(15)
(taken at $x_{0}=0$), for even $D$ the corrections of order $\mu ,\mu^2 ,\cdots$
to the energy (30) cannot be interpreted in the way like (30). On the other
hand, for odd $D\ge 3$ the energy (30) is infinite. Now, the depending on $\mu$
"corrections" find finite interpretation, but being summed up, they lead to
logarithmic divergence in the Neumann limit $\mu\rightarrow\infty$.

The analytical structure of the surface singularities of the vacuum EMT (26)
with respect to dimension $D$ is just that property
which makes dimensional regularization applicable to define
finite density (27). In view of the absence of translational symmetry
it seems quite natural for density like (27) to exist. At that time finite
renormalization of the type suggested in \cite{BEKL,AW}
eliminates the energy of the walls from $\wt E_{vac}=\wt
E_{C}(l,0)+2\wt E_w$. This procedure needs in explaination how to interpret the
expressions like (27). It should be mentioned in addition that the energy density
(27) (but not (26)) vanishes when $m=0$. This disappearance was assumed to be
characteristic of conformally invariant models \cite{BD,DW5}.

\bc
5. {\bf Conclusion}
\ec

Below some comments on literature related to the present topic are given. The
massless case for the scalar model at hand (dimension $D=2,3$) was considered in
 works \cite{MT5,BCS} and formulas consistent with (9) ( $D=3,m=0$,
\cite{BCS}) and (14) ( $D=2, m=0$, \cite{MT5}) were derived. The work \cite{BCS}
exploits an idea \cite{Sy} of replacement of the boundary with a singular
non-locally regularized potential, that occured to be equivalent to taking
potential energy (2) into account.

Formula (2.18) from \cite{AW} corresponds to our formula (31).
Doubled energy (30) appears in \cite{AW} as well, but it was not
associated there with the vacuum energy of half-space. Our formula (9) at
$\delta=+0$ (or, equivalently, (31) and (12)) gives the expression
$$
\begin{array}{c} {\ds \<\wt T_{11}\>_{0,l}=\left.
\frac{m^{D}(1-D)}{(4\pi)^{\frac{D-1}{2}}
\Gamma\(\frac{1+D}{2}\)}\;\int\limits_{1}^{\infty}\;
 \frac{\xi^2(\xi^2-1)^{\frac{D-3}{2}}}{e^{2ml\xi}-1}\; d\xi
\right|_{ml\gg 1}=} \\  \\ =
{\ds -\frac{m^{D}}{(4\pi ml)^{\frac{D-1}{2}}}\;
e^{-2ml}\;\[1+\frac{(D-1)(D+5)}{16ml}+\cdots \]},
\end{array}
\eqno{(32)}
$$
which does not coincide with formula (2.13) from \cite{Aq}. The latter, being
taken at zero temperature, includes physically unacceptable dependence of
pressure on (arbitrary) renormalization parameter. Such dependence, as it was
noted above, is characteristic of curved boundary under the condition of
coincidence between the parameters determining its curvature and size.

The author would like to acknowledge useful discussions with A.I.~Nikishov and
V.I.~Ritus and financial support from RFFR (grants 95-02-04219-a and
96-15-96463).

\newpage
\bc
{\bf References}
\ec

\end{document}